\documentclass[10pt,twocolumn,twoside]{IEEEtran}

\usepackage{multirow}
\usepackage{graphicx}
\usepackage[cmex10]{amsmath}
\usepackage{amssymb}
\newcounter{MYtempeqncnt}

\begin{document}

\title{Performance Analysis of the Ordered V-BLAST Approach over Nakagami-\emph{m} Fading Channels}

\author{Nikolaos~I.~Miridakis and Dimitrios~D.~Vergados,~\IEEEmembership{Senior Member,~IEEE}
        % <-this % stops a space
\thanks{N. I. Miridakis and D. D. Vergados are with the Department of Informatics, University of Piraeus, GR-185 34, Piraeus, Greece (e-mail: nikozm@unipi.gr, vergados@unipi.gr).}}

\markboth{}%
{Performance Analysis of the V-BLAST Approach over Nakagami-\emph{m} Fading Channels}

\maketitle

\begin{abstract}
The performance of the V-BLAST approach, which utilizes successive interference cancellation (SIC) with optimal ordering, over independent Nakagami-\emph{m} fading channels is studied. Systems with two transmit and $n$ receive antennas are employed whereas the potential erroneous decision of SIC is also considered. In particular, tight closed-form bound expressions are derived in terms of the average symbol error rate (ASER) and the outage probability, in case of binary and rectangular $\mathcal{M}$-ary constellation alphabets. The mathematical analysis is accompanied with selected performance evaluation and numerical results, which demonstrate the usefulness of the proposed approach.
\end{abstract}


\IEEEpeerreviewmaketitle

\section{Introduction}
\IEEEPARstart{T}{he} V-BLAST approach represents a cornerstone reception strategy for multiple input-multiple output (MIMO) infrastructures because it achieves a high spectral efficiency and a substantial capacity gain \cite{ref11,ref12}. It utilizes successive interference cancellation (SIC) in a number of consecutive stages. The symbol detection and the corresponding decoding at a given SIC stage can be implemented according to an optimal symbol ordering, based on the highest signal-to-noise ratio (SNR) level, or without ordering. Since SIC is quite a complex process, its average symbol error rate (ASER) performance has been studied mainly numerically (e.g. Monte Carlo simulations) and/or semi-analytically with respect to the instantaneous symbol error rate (SER).

Thereby, analytical research studies for the V-BLAST (or SIC) approach are very limited in the bibliography so far. More specifically, Loyka $et\:al$ performed an analytical framework with respect to ASER for $2\times n$ MIMO systems with optimal ordering in \cite{ref7} and for the generalized $l\times n$ case without optimal ordering in \cite{ref12}, where $l$ and $n$ denote the number of transmit and receive antennas, respectively. Nevertheless, these contributions assumed an error-free SIC approach and Rayleigh channel fading conditions.

Nakagami-\emph{m} is a versatile model, which includes the Rayleigh fading condition as a special case. To this end, an analytical framework for $2\times n$ MIMO SIC-enabled systems with optimal ordering over Nakagami-\emph{m} fading channels is presented into this letter. The merits of the proposed approach are twofold: 1) tight closed-form bound formulae for V-BLAST systems over spatially independent Nakagami-\emph{m} fading channels are derived in terms of ASER and the outage probability, thereby generalize some of the results given in \cite{ref7,ref13}; 2) a novel analytical expression for the potential error propagation of the SIC process is presented. 

\section{Statistics of the SIC Stages}
Consider a $2\times n$ MIMO SIC-enabled system with two transmit and $n\geq 2$ receive antennas. The following standard baseband discrete-time system model is employed, expressed as $\textbf{r}=\textbf{H\;s}+\textbf{w}$, where $\textbf{H}=[\textbf{h}_{1},\textbf{h}_{2}]$ denotes the $n \times 2$ channel matrix, $\textbf{r}=[r_{1},...,r_{n}]^{T}$, $\textbf{s}=[s_{1},s_{2}]^{T}$, $\textbf{w}=[w_{1},...,w_{n}]^{T}$ are the received, the transmit and the additive white Gaussian noise vector, respectively. Moreover, $\textbf{h}_{i}$ represents the \textit{i}th channel $n \times 1$ column vector, $i=1,2$ and $(.)^{T}$ denotes vector transposition.

Let $x$ be the received instantaneous SNR. The probability density function (PDF) and the cumulative distribution function (CDF) of $x$ over spatially independent Nakagami-\emph{m} fading channels are, respectively, expressed as $f_{x}(x)=\frac{\left(m/\Omega\right)^{m}}{\Gamma(m)}x^{m-1}\;\textrm{exp}\left(-\frac{m}{\Omega}x\right)$ and $\mathcal{F}_{x}(x)=\gamma\left(m,\frac{m}{\Omega}x\right)/\Gamma(m)$, where $\Gamma(.)$ is the gamma function \cite[eq. 8.310/1]{ref1}, $\gamma(.,.)$ is the lower incomplete gamma function \cite[eq. 8.350/1]{ref1}, $m=2n\times m_{N}$ is the normalized Nakagami-\emph{m} fading parameter with respect to the number of transmit and receive antennas (the factor $2$ indicates the number of transmit antennas), $m_{N}\geq \frac{1}{2}$ is the distribution shape parameter (which indicates the fading severity), $\Omega=\mathcal{E}[x]$ is the average signal power and $\mathcal{E}[.]$ denotes expectation.

The reception process is implemented successively in a number of stages, proportional to the number of the transmit antennas. The key idea is that as a given symbol is detected, decoded and then canceled from the composite signal at the \textit{i}th SIC stage, the remaining signal at the next stage experiences better channel conditions in terms of SNR and, hence, a better ASER performance. For a detailed description of the SIC architecture and methodology, see \cite{ref10}.

\subsection{First SIC Stage}
The upper bound on the CDF of $x$ given at the first SIC stage, $F_{\textbf{1}}(x)$, is obtained as \cite[eq. 5]{ref13}
\begin{align}
\nonumber
&F_{\textbf{1}}(x)=\\
&(n-1)\int^{1}_{0}\mathcal{F}_{x}\left(\frac{x}{t}\right)t^{n-2} dt\geq (n-1)\int^{1}_{0}\mathcal{F}^{2}_{x}\left(\frac{x}{t}\right)t^{n-2} dt,
\label{F1}
\end{align}
where the expression on the right hand side of the inequality represents the actual CDF of $x$ at the first SIC stage, which is analytically infeasible to be obtained in a straightforward closed-form solution, mainly due to the involvement of $\gamma(.,.)^{2}$. However, unlike the actual CDF of $x$, the respective upper bound can be derived in a closed-form expression. Based on (\ref{F1}) while utilizing first \cite[eq. 9.31/2]{ref1} and then \cite[eq. 7.811/2]{ref1} we have that 
\begin{equation}
F_{\textbf{1}}(x)=\frac{(n-1)}{\Gamma(m)}\  G^{1,2}_{3,2}\left[\left(\frac{\Omega}{m\;x}\right)~\vline
\begin{array}{c}
2-n,1-m,1 \\
0,1-n 
\end{array} \right],
\label{F1complete}
\end{equation}
where $G^{m,n}_{p,q}[~.~\vline~.~]$ is the Meijer's \emph{G} function \cite[eq. 9.30]{ref1}.

Thereby, taking the first derivative of (\ref{F1complete}), the PDF of $x$ given at the first SIC stage is expressed as
\begin{align}
\nonumber
f_{\textbf{1}}(x)&= \frac{\partial F_{\textbf{1}}(x)}{\partial x}\\ &=\frac{(n-1)\left(\frac{m}{\Omega}\right)^{n-1}}{\Gamma(m)}\;x^{n-2}\;\Gamma\left(m-n+1,\frac{m}{\Omega}x\right),
\label{f1complete}
\end{align}
where $\Gamma(.,.)$ denotes the upper incomplete gamma function \cite[eq. 8.350/2]{ref1}.

\subsection{Second SIC Stage}
The CDF of $x$ given at the second SIC stage,  $F_{\textbf{2}}(x)$, is obtained as
\begin{equation}
F_{\textbf{2}}(x)=\textrm{min}[x_{\textbf{1}},x_{\textbf{2}}]=1-\left[1-\mathcal{F}_{x}(x)\right]^{2}= 2\: \mathcal{F}_{x}(x)-\mathcal{F}^{2}_{x}(x).
\label{F2}
\end{equation}
In fact, the outage probability at the second stage is twice the corresponding one at the first stage \cite[eq. 35]{ref7}, i.e. $2\: \mathcal{F}_{x}(2x)-\mathcal{F}^{2}_{x}(2x)$. This occurs because the post-processing noise power at the second stage is twice of the branch noise power level, i.e $\sigma^{2}_{i}=(n-2+i)\;\sigma^{2}$, where $\sigma^{2}_{i}$ and $\sigma^{2}$ denote the noise power at the \textit{i}th SIC stage and the total noise power, respectively. For a detailed description of this effect, see \cite[Appendix I]{ref7}. Thus, we have that
\begin{equation}
F_{\textbf{2}}(x)\approx 2\: \mathcal{F}_{x}(2x)=\frac{2}{\Gamma(m)}\ \gamma \left(m,\frac{2 m}{\Omega}x\right),\ \ \Omega \rightarrow   \infty.
\label{F2real}
\end{equation}
It should be noted that the approximation of (\ref{F2real}) converges to the actual $F_{\textbf{2}}(x)$, as given in (\ref{F2}), for typically medium/high average SNR values. However, the above mentioned fluctuation is maintained small even in the low SNR regime, as demonstrated by the numerical results of the next section. Hence, the PDF of $x$ given at the second SIC stage is derived as
\begin{equation}
f_{\textbf{2}}(x)=\frac{\partial F_{\textbf{2}}(x)}{\partial x}=\frac{2^{m+1}\left(\frac{m}{\Omega}\right)^{m}}{\Gamma(m)}\;x^{m-1}\;\textrm{exp}\left(-\frac{2m}{\Omega}x\right).
\label{f2complete}
\end{equation}

\section{Performance Analysis}
\subsection{ASER}
The average symbol error rate (ASER) at the \textit{i}th SIC stage, $\bar{P}_{s,i}$, can directly be evaluated by averaging the conditional error probability (CEP), $P_{e}(\epsilon|x_{i})$, over $f_{i}(x_{i})$ and is expressed as
\begin{equation}
\bar{P}_{s,i}\triangleq \int^{\infty}_{0}P_{e}(\epsilon|x_{i})\ f_{i}(x_{i})dx_{i}.
\label{Ps}
\end{equation}
In case of binary modulations, CEP is defined as \cite{ref3}
\begin{equation}
P_{e}(\epsilon|x_{i})=\frac{\Gamma(\beta,\alpha x_{i})}{2 \Gamma(\beta)},
\label{CEPB}
\end{equation}
where $\alpha$ and $\beta$ are certain constants that define the modulation type. In case of rectangular $\mathcal{M}$-ary modulations and higher values of the average input SNR, CEP is defined as \cite{ref3}
\begin{equation}
P_{e}(\epsilon|x_{i})=\alpha\; \textrm{erfc}\left(\sqrt{\beta \;x_{i}}\right),
\label{CEPM}
\end{equation}
where $\textrm{erfc}(.)$ denotes the complementary error function \cite[eq. 8.250/4]{ref1}.

The total ASER in a $2\times n$ MIMO SIC-enabled system is expressed as
\begin{equation}
\bar{P}_{s,total}=\bar{P}_{s,\textbf{1}}+\bar{P}_{s,\textbf{2}}(1-\bar{P}_{s,\textbf{1}})=\bar{P}_{s,\textbf{1}}+\bar{P}_{s,\textbf{2}}-\overline{P_{e,\textbf{1}}*P_{e,\textbf{2}}}.
\label{ASERT}
\end{equation} 
It is worth noting that (\ref{ASERT}) is rigorous and accounts for the potential error propagation of the SIC process (i.e when no error occurs at the first stage and an error occurs at the second stage or both the SIC stages are erroneous). Unlike the first two terms of (\ref{ASERT}), the third term involves a conditioning on both $x_{\textbf{1}}$ and $x_{\textbf{2}}$, which are not statistically independent. Hence, the more complicated bivariate (correlated) PDF is required in this case. Moreover, $\overline{P_{e,\textbf{1}}*P_{e,\textbf{2}}}$ represents a \emph{second order} statistic, which may fluctuate the performance of the total ASER, especially in the low SNR regions (where the presence of the error propagation is more emphatic). In the following, tight closed formulae are derived with respect to the total ASER, for binary and $\mathcal{M}$-ary modulation schemes\footnote{Note that the  average bit error rate at the \textit{i}th stage, $\bar{P}_{b,i}$, can be easily deduced from the corresponding ASER, assuming that $\bar{P}_{b,i}\cong\bar{P}_{s,i}/\textrm{log}_{2}\mathcal{M}$.}. 

\begin{figure*}[!t]

\setcounter{MYtempeqncnt}{\value{equation}}
\setcounter{equation}{10}
\footnotesize
\begin{align}
\bar{P}^{(Binary)}_{s,\textbf{1}}=
\textstyle \frac{1}{2}-\frac{(n-1)\alpha^{\beta}\Gamma\left(\beta+m\right)}{2\beta(\beta+n-1)\Gamma(\beta)\Gamma(m)\left(\frac{m}{\Omega}\right)^{\beta}} \ {}_3 F_2\left(m+\beta,\beta,n+\beta-1;\beta+1,n+\beta;-\frac{\Omega \alpha}{m}\right).
\label{ASERB}
\tag{11}
\end{align}
\setcounter{equation}{\value{MYtempeqncnt}}

\setcounter{MYtempeqncnt}{\value{equation}}
\setcounter{equation}{11}
\begin{align}
\bar{P}^{(\mathcal{M}-ary)}_{s,\textbf{1}}=\textstyle \frac{\alpha \left(n-1\right)\left(\frac{m}{\beta \Omega}\right)^{m}\Gamma\left(m+\frac{1}{2}\right)}{\sqrt{\pi}\;\Gamma(m) m (m-n+1)}\ {}_3F_2\left(m-n+1,m,m+\frac{1}{2};m-n+2,m+1;-\frac{m}{\beta \Omega}\right)-\frac{\alpha \;\left(\frac{m}{\beta \Omega}\right)^{n-1}\Gamma(m+n-1) \Gamma\left(n-\frac{1}{2}\right)}{\sqrt{\pi} \Gamma(m)}.
\label{ASERM}
\tag{12}
\end{align}
\setcounter{equation}{\value{MYtempeqncnt}}
\hrulefill
%\vspace*{4pt}

\setcounter{MYtempeqncnt}{\value{equation}}
\setcounter{equation}{12}
\begin{align}
\nonumber
\bar{P}^{(Binary)}_{e,\textbf{2}}=\textstyle 1-\frac{\alpha^{\beta}\;\Gamma\left(m+\beta\right)}{\beta\;\Gamma(\beta)\Gamma(m)\left(\frac{2 m}{\Omega}\right)^{\beta}} \ {}_2 F_1\left(\beta,m+\beta;\beta+1;-\frac{\alpha \Omega}{2 m}\right).
\label{BER2b}
\tag{13}
\end{align}
\setcounter{equation}{\value{MYtempeqncnt}}

\setcounter{MYtempeqncnt}{\value{equation}}
\setcounter{equation}{13}
\begin{align}
\nonumber
\bar{P}^{(\mathcal{M}-ary)}_{e,\textbf{2}}=\textstyle 2 \alpha-\frac{4\;\alpha\;\sqrt{\beta}\; \Gamma\left(m+\frac{1}{2}\right)}{\Gamma(m)\sqrt{\frac{2 m}{\Omega}}\sqrt{\pi}}\ {}_2 F_1\left(\frac{1}{2},m+\frac{1}{2};\frac{3}{2};-\frac{\beta \Omega}{2 m}\right).
\label{BER2m}
\tag{14}
\end{align}
\setcounter{equation}{\value{MYtempeqncnt}}
\hrulefill
%\vspace*{4pt}

\setcounter{MYtempeqncnt}{\value{equation}}
\setcounter{equation}{20}
\begin{align}
\nonumber
\overline{P_{e,\textbf{1}}*P_{e,\textbf{2}}}^{(Binary)}&=\textstyle \frac{1}{4^{m+\frac{1}{2}}\Gamma(m)\Omega^{m+1}(1-\rho)\rho^{\frac{m-1}{2}}\Gamma(\beta)}\displaystyle \sum^{\infty}_{k=0}\textstyle {\frac{\sqrt{\rho}}{4\alpha^{k+m}(1-\rho)\;\Omega\; \Gamma(k+m)k!}}\displaystyle \sum^{\infty}_{j=0}\textstyle \frac{\Gamma(\beta+m+k+j)}{\Gamma^{2}(j+1)2(k+j+m)}\\
&\textstyle \times \left(\frac{1}{2\sqrt{\alpha }\;\Omega(1-\rho)}\right)^{2j}\Bigg[2\psi(j+1)+\frac{1}{m+k+j}-\psi(\beta+m+k+j)-2\;\textrm{ln}\left(\frac{1}{2\sqrt{\alpha}(1-\rho)\Omega}\right)\Bigg].
\label{BER1,2b}
\tag{19}
\end{align}
\setcounter{equation}{\value{MYtempeqncnt}}

\setcounter{MYtempeqncnt}{\value{equation}}
\setcounter{equation}{21}
\begin{align}
\nonumber
\overline{P_{e,\textbf{1}}*P_{e,\textbf{2}}}^{(\mathcal{M}-ary)}&=\textstyle \frac{\alpha}{4^{m}\Gamma(m)\Omega^{m+1}(1-\rho)\rho^{\frac{m-1}{2}}}\displaystyle \sum^{\infty}_{k=0}\textstyle {\frac{\sqrt{\rho}}{2\beta^{k+m}\sqrt{\pi}(1-\rho)\;\Omega \;\Gamma(k+m)k!}}\displaystyle \sum^{\infty}_{j=0}\textstyle \frac{\Gamma\left(k+j+m+\frac{1}{2}\right)}{\Gamma^{2}(j+1)2(k+j+m)}\\
&\textstyle \times \left(\frac{1}{2\sqrt{\beta }\;\Omega(1-\rho)}\right)^{2j}\Bigg[2\psi(j+1)+\frac{1}{m+k+j}-\psi\left(m+k+j+\frac{1}{2}\right)-2\;\textrm{ln}\left(\frac{1}{2\sqrt{\beta }(1-\rho)\Omega}\right)\Bigg].
\label{BER1,2m}
\tag{20}
\end{align}
\setcounter{equation}{\value{MYtempeqncnt}}
\hrulefill
%\vspace*{4pt}

\end{figure*}

\subsubsection{ASER at the First SIC Stage}
Based on (\ref{f1complete}) and (\ref{CEPB}) while utilizing \cite[eq. 2.10.6/1]{ref2}, the ASER for binary modulations is derived, as given in (\ref{ASERB}), where ${}_pF_q$ denotes the generalized hypergeometric function \cite[eq. 9.14/1]{ref1}. Similarly, based on (\ref{f1complete}) and (\ref{CEPM}), the ASER for $\mathcal{M}$-ary modulations is obtained in (\ref{ASERM}), by invoking \cite[eq. 2.10.8/2]{ref2}.

\subsubsection{ASER at the Second SIC Stage}
In case of binary modulations, based on (\ref{f2complete}) and (\ref{CEPB}) while utilizing \cite[eq. 2.10.3/2]{ref2}, the corresponding ASER is obtained, as given in (\ref{BER2b}). In case of $\mathcal{M}$-ary modulations, based on (\ref{f2complete}) and (\ref{CEPM}) while invoking \cite[eq. 2.8.5/6]{ref2} and after performing some straightforward algebraic manipulations, the corresponding ASER is derived in (\ref{BER2m}). 

\subsubsection{Cross-Product (Correlated) ASER}
The statistically correlated cross-product term can be obtained by averaging CEP over the PDF of such an event and is defined as
\setcounter{equation}{14}
\begin{equation}
\overline{P_{e,\textbf{1}}*P_{e,\textbf{2}}}=\int^{\infty}_{0} P_{e}(\epsilon|y)\ f_{y}(y)dy,
\label{CROSS}
\end{equation}
where $y=x_{\textbf{1}} x_{\textbf{2}}$, $P_{e}(\epsilon|y)$ denotes the CEP on $y$ and $f_{y}(y)$ is the cross-product PDF, since (\ref{F2real}) and the following condition hold \cite[eq. 6.74]{ref4}
\begin{equation}
f_{y}(y)=\int^{\infty}_{0}\frac{1}{x_{\textbf{1}}}~f_{x_{\textbf{1}},x_{\textbf{2}}}\left(x_{\textbf{1}},\frac{y}{x_{\textbf{1}}}\right)dx_{\textbf{1}}.
\label{CROSSPDF}
\end{equation}
The bivariate Nakagami-\emph{m} PDF\footnote[2]{It is assumed that $\Omega_{\textbf{1}}=\Omega_{\textbf{2}}=\Omega$, i.e. the pre-processing average SNR is identical for all branches and for both transmitters.}, $f_{x_{\textbf{1}},x_{\textbf{2}}}(x_{\textbf{1}},x_{\textbf{2}})$, is expressed as \cite[eq. 2]{ref9}
\begin{align}
\nonumber
f_{x_{\textbf{1}},x_{\textbf{2}}}(x_{\textbf{1}},x_{\textbf{2}})&=\frac{4(x_{\textbf{1}}x_{\textbf{2}})^{\frac{m}{2}}\;\textrm{exp}\left(-\frac{x_{\textbf{1}}+x_{\textbf{2}}}{\Omega(1-\rho)}\right)}{\Gamma(m)\Omega^{m+1}(1-\rho)\rho^{\frac{m-1}{2}}}\\
&\times \mathcal{I}_{m-1}\left(\frac{2\sqrt{\rho x_{\textbf{1}} x_{\textbf{2}}}}{\Omega(1-\rho)}\right),
\label{jointf}
\end{align}  
where $\mathcal{I}_{\nu}(.)$ denotes the modified Bessel function of the first kind and order $\nu$ \cite[Appendix II.10/1]{ref2} and $\rho$ represents the correlation coefficient, which is denoted as $\rho=\textrm{cov}(x_{\textbf{1}},x_{\textbf{2}})/\sqrt{\textrm{var}(x_{\textbf{1}})\textrm{var}(x_{\textbf{2}})}$, where $\textrm{var}(.)$ and $\textrm{cov}(.,.)$ denote variance and covariance, respectively. By invoking \cite[eq. 3.478/4]{ref1}, (\ref{CROSSPDF}) can be easily resolved as
\begin{align}
\nonumber
f_{y}(y)&=\frac{y^{\frac{m}{2}}}{4^{m}\Gamma(m)\Omega^{m+1}(1-\rho)\rho^{\frac{m-1}{2}}}\\
&\times \mathcal{I}_{m-1}\left(\frac{\sqrt{\rho y}}{\Omega(1-\rho)}\right)\mathcal{K}_{0}\left(\frac{\sqrt{y}}{\Omega(1-\rho)^{2}}\right),
\label{FY}
\end{align}
where $\mathcal{K}_{\nu}(.)$ denotes the modified Bessel function of the second kind and order $\nu$ \cite[eq. II.10/2]{ref2}.

Unfortunately, substituting (\ref{FY}) and (\ref{CEPB}) or (\ref{CEPM}) in (\ref{CROSS}) does not provide a straightforward tabulated closed-form solution. Nonetheless, by applying the infinite series representation of $\mathcal{I}_{\nu}(.)$ \cite[Appendix II.10/1]{ref2} in (\ref{FY}), while utilizing \cite[eq. 2.16.1/31]{ref2}, the correlated ASER for binary modulations is obtained in (\ref{BER1,2b}), where $\psi(.)$ denotes the digamma function \cite[eq. 8.360/1]{ref1}. In addition, by using (\ref{CEPM}) and (\ref{FY}) in (\ref{CROSS}) and with the aid of \cite[eq. 2.16.1/25]{ref2}, the corresponding ASER for $\mathcal{M}$-ary modulations is derived in (\ref{BER1,2m}). Note that the summation terms in (\ref{BER1,2b}) and (\ref{BER1,2m}) converge very rapidly for various system scenarios (e.g. (\ref{BER1,2m}) requires only $4$ summation terms in order to converge up to the $9^{\textrm{th}}$ digit, when $m=2$ and $\rho=0.7$).

\begin{figure}[!ht]
\centering
\includegraphics[keepaspectratio,width=2.6in]{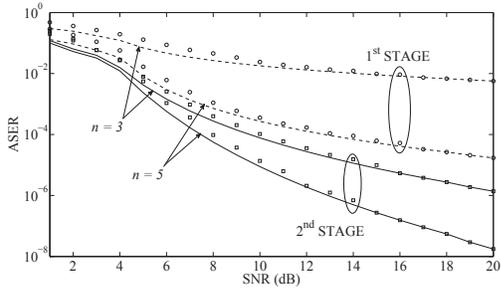}
\caption{The ASER of the first and the second SIC stage vs. different values of the average input SNR, for a $2 \times n$ MIMO system, when $m_{N}=1$ and a BPSK modulation scheme is considered. The numerical and the analytical results are indicated by marker signs and lines, respectively.}
\label{fig.1}
\end{figure}

\begin{figure}[!ht]
\centering
\includegraphics[keepaspectratio,width=2.6in]{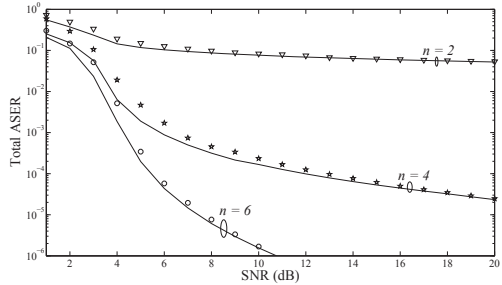}
\caption{The total ASER, as given in (\ref{ASERT}), vs. different modulation schemes and several values of the average input SNR, for a $2 \times n$ MIMO system, when $m_{N}=2$. The numerical and the analytical results are indicated by marker signs and solid lines, respectively.}
\label{fig.2}
\end{figure}

\begin{figure}[!ht]
\centering
\includegraphics[keepaspectratio,width=2.6in]{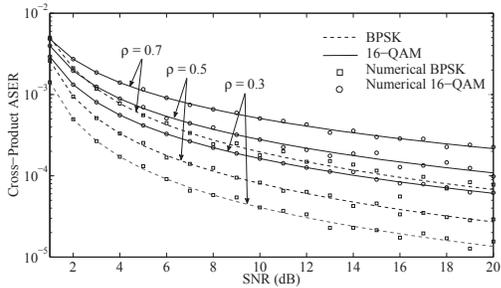}
\caption{The cross-product ASER vs. different values of the average input SNR, for a $2 \times 2$ MIMO system, under various correlation conditions and different modulation schemes, when $m_{N}=0.5$.}
\label{fig.3}
\end{figure}

Figs.~\ref{fig.1} and~\ref{fig.2} show the ASER performance of the two consecutive SIC stages and the total ASER, respectively, in various system scenarios. It is obvious that the corresponding ASER is sharply affected with an increase of the number of receive antennas and with a reduction of the channel fading severity (e.g. in higher $m_{N}$ values). Moreover, in the worst system scenario in terms of ASER, where only two receive antennas are employed and the channel fading severity is quite intense ($m_{N}=0.5$), the influence of the cross-product ASER is shown in Fig.~\ref{fig.3}. The numerical results at the above mentioned configurations have obtained via numerical evaluation based on the actual statistics, as given at the right hand side of (\ref{F1}) and (\ref{F2}), and then by utilizing (\ref{ASERT}). A slight difference on the ASER performance between the analytical bound formulations and the respective exact numerical verification is observed whereas quite an effective computational gain is achieved by performing the proposed approach.

\subsection{Outage Probability}
\begin{figure}[!t]
\centering
\includegraphics[keepaspectratio,width=2.6in]{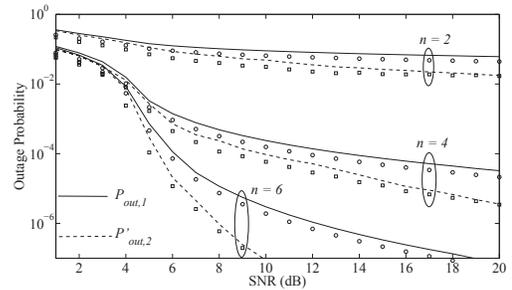}
\caption{The outage probability vs. different values of the average input SNR for a $2\times n$ MIMO system, when $m_{N}=2$ and a BPSK modulation scheme is considered. The numerical and the analytical results are indicated by marker signs and lines, respectively.}
\label{fig.4}
\end{figure}

The outage probability, $P_{out,\textbf{1}}(x_{th})$, at the first SIC stage is directly obtained from (\ref{F1complete}), where $x_{th}$ denotes a threshold SNR value. The outage probability at the second stage, $P_{out,\textbf{2}}(x_{th})$, conditioned on an error-free first stage is given in (\ref{F2real}). The corresponding unconditional outage probability at the second stage, $P_{out,\textbf{2}}'(x_{th})$, which considers the potential erroneous decision at the first stage, is expressed as $P_{out,\textbf{2}}'(x_{th})=F_{\textbf{2}}(x_{th})(1-\bar{P}_{s,\textbf{1}})+\bar{P}_{s,\textbf{1}}$. Note, that $P_{out,\textbf{1}}$ is independent of the error propagation whereas $P_{out,\textbf{2}}'$ is typically upper bounded by $\bar{P}_{s,\textbf{1}}$. Fig. \ref{fig.4} indicates the outage probability in various average SNR regions for $2\times n$ MIMO systems. It is obvious that as the spatial diversity gain increases, the outage performance improves (i.e. $P_{out}$ decreases) for both SIC stages. 
   
\section{Performance Bounds for the Generalized V-BLAST}
Consider an $l \times n$ MIMO system with $l\leq n$ transmit antennas. Hence, $\textbf{H}=[\textbf{h}_{1},...,\textbf{h}_{l}]$, $\textbf{r}=[r_{1},...,r_{n}]^{T}$, $\textbf{s}=[s_{1},...,s_{l}]^{T}$, $\textbf{w}=[w_{1},...,w_{n}]^{T}$ while SIC is implemented in $l$ consecutive stages. In this case, the distribution of SNR can not be resolved in a closed-form expression for the \textit{i}th stage, when $i>1$ (even if the signal undergoes Rayleigh fading \cite[eq. 25]{ref19}).
%Unfortunately, a tractable closed-form solution for the PDF/CDF of each stage is not feasible in this case (not even in the simpler Rayleigh fading scenario). This occurs mainly due to the angle between $\textbf{h}_{i}$ and the sub-space spanned all the other column vectors at the \textit{i}th stage, where the corresponding PDF of this angle can not be analytically resolved. To this end, 
Fortunately, there is an upper bound expression for the outage of the first SIC stage, given as \cite{ref13}
\begin{align}
\setcounter{equation}{20}
\textstyle F_{\textbf{1}}^{(l \times n)}(x)=\binom{n-1}{l-1}(l-1) \displaystyle \int^{1}_{0}\textstyle \left[\mathcal{F}_{x}\left(\frac{x}{t}\right)\right]^{l}t^{n-l}(1-t)^{l-2} dt,
\label{general}
\end{align}
which represents a generalization of (\ref{F1}). Note that (\ref{general}) can not be evaluated in a closed formulation mainly due to the involvement of $\gamma(.,.)$ within $F_{x}(x)$. However, comparing (\ref{F1}) and (\ref{general}) whereas recognizing that the diversity gain of an $l \times n$ V-BLAST at the first stage (which is $n-l+1$) is always lower than a $2 \times n$ V-BLAST, i.e. $n-l+1<n-1$ for $l>2$, we have that $F_{\textbf{1}}^{(l \times n)}\geq F_{\textbf{1}}^{((l-1) \times n)}\geq...\geq F_{\textbf{1}}^{(2 \times n)}$, while equiprobably it holds that $\bar{P}^{(l \times n)}_{s,\textbf{1}}\geq \bar{P}^{((l-1) \times n)}_{s,\textbf{1}}\geq...\geq \bar{P}^{(2 \times n)}_{s,\textbf{1}}$. Thereby, (\ref{F1complete}) and (\ref{ASERB}) (or (\ref{ASERM})) can serve as sharp closed-form lower bounds for the generalized $l \times n$ case with respect to the outage performance and ASER, respectively.
\ifCLASSOPTIONcaptionsoff
  \newpage
\fi

\end{document}